\documentclass[5p]{elsarticle}
\usepackage{lineno,hyperref}
\usepackage{cleveref}
\usepackage{graphicx} 
\usepackage{subcaption}
\usepackage{ulem}
\usepackage{booktabs}
\usepackage{algorithm,algpseudocode}
\usepackage{amsmath}
\usepackage{tabularx}
\usepackage{multirow}
\usepackage{rotating}
\usepackage{pdflscape}
\usepackage{fancyhdr}
\usepackage{textcomp}

\modulolinenumbers[5]

\begin{document}
	
\captionsetup[table]{
	labelsep = newline,
	labelfont = bf,
	name = Table,
	justification=justified,
	singlelinecheck=false,
	skip = \medskipamount}	
	
	\newcolumntype{L}[1]{>{\raggedright\arraybackslash}p{#1}}
	\newcolumntype{C}[1]{>{\centering\arraybackslash}p{#1}}
	\newcolumntype{R}[1]{>{\raggedleft\arraybackslash}p{#1}}

\begin{frontmatter}
		
\title{A hybrid automated detection of epileptic seizures in EEG based on wavelet and machine learning techniques}
		
		

\author[mymainaddress,mysecondaryaddress]{Asmaa Hamad}
\author[Add2,mysecondaryaddress]{Aboul Ella Hassanien}
\author[Add2]{Aly A. Fahmy}	
\author[mymainaddress]{Essam H. Houssein}
	
\address[mymainaddress]{Faculty of Computers and Information, Minia University, Egypt}		
\address[Add2]{Faculty of Computers and Information, Cairo University, Egypt}
\address[mysecondaryaddress]{Scientific Research Group in Egypt (SRGE) http://www.egyptscience.net}

\begin{abstract}
Epilepsy is a neurological condition such that it affects the brain and the nervous system. It is characterized by recurrent seizures, which are physical reactions to sudden, usually brief, excessive electrical discharges in a group of brain cells. Hence, seizure identification has great importance in clinical therapy of epileptic patients. Electroencephalogram (EEG) is one of the main biomarker that can measure voltage fluctuations of the brain and EEG data analysis helps to investigate the patient with epilepsy syndrome as epilepsy leaves their signature in EEG signals. In this paper, the Discrete Wavelet Transform (DWT) is applied to EEG signals to pre-processing, decompose it till the 4th level of decomposition tree.Various features like Entropy, Min, Max, Mean, Median, Standard deviation, Variance, Skewness, Energy and Relative Wave Energy (RWE) were computed in terms of detailed coefficients and the approximation coefficients of the last decomposition level.Then, the extracted features are evaluated by three modern machine-learning classifiers such as Radial Basis Function based Support Vector Machine (SVMRBF), k-Nearest Neighbor (KNN) and Naive Bayes (NB). The experimental results demonstrate that the highest classification accuracy (100\%) for normal subject data versus epileptic data is obtained by SVMRBF. the corresponding accuracy between normal subject data and epileptic data using KNN and NB is obtained as 99.50\% and 99\% for the eyes open and eyes closed conditions, respectively. The similar accuracies, while comparing the interictal and ictal data, are obtained as 99\%, 97.50\% and 98.50\% using the SVMRBF, KNN and NB classifiers, respectively. These accuracies are quite higher than earlier results published.
\end{abstract}

\end{frontmatter}

\section{Introduction}
\label{Sec:Introduction}
Epilepsy is the second most widespread neurological condition after Alzheimer's disease and stroke visible in primary practice worldwide with an approximate prevalence of 5.8 per 1000 population in the advanced world and between 10.3 per 1000 to 15.4 per 1000 in developing countries \cite{josephson2016systematic}. People with epilepsy are two or three times more likely to die prematurely when compared to a normal person. Therefore, diagnosing and predicting epileptic seizures precisely appear to be particularly important, which is able to fetch more effective prevention and treatment for the patients. Clinically to predict and diagnose epileptic seizures, the brain activities are to be observed through EEG signals which contain the markers of epilepsy. Electroencephalography (EEG) is the recording of the electrical activity of the brain, regularly taken through several electrodes at the scalp. EEG contains lots of worthy information relating to the numerous physiological states of the brain and thus is a very useful tool for understanding the brain disease, such as epilepsy \cite{C2_39_guo2011automatic}. EEG signals of epileptic patients exhibit two states of abnormal activities namely interictal or seizure free (in-between epileptic seizures) and ictal (in the course of an epileptic seizure) \cite{acharya2015application}. The interictal EEG signals are transitory waveforms and exhibit spikes, sharp or spiky waves. The ictal EEG signals are persistent waveforms with spikes and sharp wave complexes. Epilepsy can be revealed by conventional methods by well-trained and experienced neurophysiologists by visual inspection of long durations of EEG signals, this is time-consuming, tedious and individual. Hence, in order to overcome these limitations, a computer-aided detection (CAD) of epileptic EEG signals can be utilized.

The EEG signals are ordinarily decomposed into five sub-bands: delta (0-4 Hz), theta (4-8 Hz), alpha (8-12 Hz), beta (13-30 Hz), and gamma (30-60 Hz) \cite{w1,w2}. Alpha waves are rhythmic and its amplitude is low. Each region of the brain has the distinguishing marks of alpha rhythm but mostly it is recorded from the occipital and parietal zone. It oscillates from an adult in the awake and relaxed situation with eyes closed. Beta waves are irregular and its amplitude is very low. It is primarily recorded from temporal and frontal lobe. It oscillates from during the deep sleep, mental activity and is related with remembering. Theta waves are rhythmic and its amplitude is low-medium. It oscillates from the children in a sleep state, drowsy adult, and sentimental distress occipital lobe. Delta waves are slow and its amplitude is high. It varies from adult and normal sleep rhythm. Gamma waves are the rapid brainwave frequency with the smallest amplitude \cite{kalaivani2014analysis}.

 Several algorithms have been developed in the literature to improve the detection and classification of epileptic EEG Signals. In \cite{C2_36_nicolaou2012detection}, the authors proposed automated epileptic seizure detection based on SVM to classify segments of normal and epileptic, permutation entropy (PE) is used as a feature.  This work achieved a total averaged accuracy of 85.16\%. Also in \cite{C2_15_guo2009classification}, the authors employed the wavelet transform and the relative energy as input features to an Artificial Neural Network to classify normal and epileptic EEG Signal. A maximum accuracy of 95.2\% is achieved. In \cite{C2_39_guo2011automatic}, the authors applied genetic programming (GP) for automatic feature extraction to improve  both of the K-nearest neighbor (KNN) performance and the feature dimension reduction to classify healthy subjects with eyes open versus ictal EEG segment and also to classify interictal EEG records with ictal records versus EEG recorded from healthy subjects with eye open, a maximum accuracy of 99.2\% was obtained. In addition, in \cite{C2_40_kumar2014epileptic}, the authors developed a scheme for detecting epileptic seizures from EEG data recorded from epileptic patients and normal subjects. This scheme is based on DWT analysis and approximate entropy (ApEn) of EEG signals. SVM and (feed-forward backpropagation neural network) FBNN are used for classification purpose. This work achieved a total averaged accuracy of 95.92\%.

Moreover, in \cite{C2_41_li2011clustering}, the authors presented a clustering-based least square support vector machine approach for the classification of EEG signals. Their proposed approach comprises the following two stages. In the first stage, clustering technique (CT) has been used to extract representative features of EEG data. In the second stage, least square support vector machine (LS-SVM) is applied to the extracted features to classify EEG signals, a total averaged accuracy of 94.18\% was obtained. Also, in \cite{C2_42_wang2011best}, the authors proposed a hierarchical epileptic seizure detection approach that classifies healthy subjects with eyes open versus ictal segments. In this approach, the original EEG signals performed by wavelet packet coefficients and using basis-based wavelet packet entropy method to extract feature. In the training stage, hybrid the K-Nearest Neighbour (KNN) with the cross-validation (CV) methods are utilized and achieved 99.449\% of accuracy. In addition, in \cite{C2_43_kumar2014epilepticfuzzy}, the authors presented an approach to classify EEG signals into healthy/interictal versus ictal EEGs using fuzzy approximate entropy (fApEn). In their approach, support vector machine (SVM) with RBF is utilized for classification purpose. Their work achieved a total averaged accuracy of 98.457\%. Also, in \cite{C2_44_acharya2012automated}, the authors combined a mixture of entropy measure like Approximate Entropy (ApEn), Sample Entropy (SampEn) and Phase Entropy and Fuzzy Logic Classifier. A total accuracy of 98.1\% was achieved. Furthermore, in \cite{Rpaper2017}, authors proposed  entropy measure Q-based K-NN entropy by compute K-NN entropy at different frequency scales of the EEG signa for the classification of seizure, seizure-free and normal EEG signals. A total average accuracy of 99.2\% was achieved.

In \cite{C2_45_supriya2016weighted}, the authors proposed an epileptic seizure detection technique from brain EEG signals. The EEG time series are transformed into a weighted visibility graph (WVG). The modularity and average weighted degree are extracted based on WVG. Moreover, SVM and KNN are utilized  to classify EEG signals  into healthy/interictal versus ictal EEGs with a total averaged accuracy of 94.94\%. Also in \cite{C2_46_guo2010epileptic}, the authors presented an automatic epileptic seizure detection method, which uses approximately entropy features derived from multiwavelet transform. Artificial neural network (ANN) is combined with entropy to classify healthy subjects with eyes open versus ictal EEGs and also to classify interictal EEG records and healthy subjects versus ictal EEGs; a maximum accuracy of 99.85\% was obtained. In the same manner, in \cite{C2_4_chua2008automatic}, the authors attempted to classify healthy subjects with eyes open versus ictal EEG segment and also classify interictal EEG records versus the ictal records. Their work was based on higher order spectra and power spectral density combined with Gaussian classifier. It resulted in 93.11\% of performance accuracy. Finally, in \cite{C2_47_mirzaei2011statistical}, the authors classified healthy subjects versus ictal segments using spectral entropy (SpecEn). No accuracy was declared, only a T-student statistical test has been conducted.

In this paper, each EEG signal is decomposed into five constituent EEG sub-bands by DWT. DWT is used for time-frequency analysis giving a quantitative evaluation of numerous frequency bands of clinical brain wave. The EEG epochs were analyzed into various frequency bands by using fourth-order Daubechies (db4) wavelet function up to 4th-level of the decomposition. The statistical parameter like entropy, min, max, mean, median, standard deviation, variance, Skewness,  energy and Relative Wave Energy (RWE) were computed for feature extraction and classification experiments are performed on different EEG dataset by using three most popular machine learning classifiers named as, SVMRBF, KNN and NB. The experimental results are quite promising with 100\% accuracy in the classification of EEG signals of epileptic seizure activity set (E) and healthy person (A and B). Moreover, the results for other classification test cases also suggest that our proposed technique is best appropriate to differentiate between different kinds of EEG signals.

The remainder this paper is organized as follows: Section \ref{Materials and Methods} will introduce the materials and methods. In Section \ref{model} The proposed classification model for EEG including pre-processing of EEG, method for extracting features from EEG signals and classification are described. In Section \ref{Results} experimental results and discussions are introduced. Finally, in Section \ref{Sec:Conclusion} the conclusion and future work are presented.

\section{Materials and Methods} \label{Materials and Methods}
This section introduces the material and methods used in this paper.
\subsection{Description of EEG Dataset}
\label{Sec:EEG}
The data utilized in this paper was taken from publicly available data at the Department of Epileptology, University of Bonn (\cite{dataset}). This dataset includes five sets (denoted as A, B, C, D and E), each including 100 single-channel EEG segments of 23.6 sec duration, with a sampling rate of 173.6 Hz. Where, each data segment contains N=4097 data points accumulated at intervals of 1/173.61th of 1s.These segments were chosen and cut out from continuous multichannel EEG recordings after a visual investigation for artifacts, e.g., due to muscle activity or eye movements. The data sets A and B comprised from segments taken from surface EEG recordings that were carried out on five healthy volunteers using a unified electrode placement scheme. The volunteers were relaxed in an awake state with eyes open (A) and eyes closed (B), respectively. The data sets C, D, and E are recorded from the epileptic subjects through intracranial electrodes for interictal and ictal epileptic activities. All of the subjects had achieved complete seizure control after resection of one of the hippocampal formations, which was therefore accurately diagnosed to be the epileptogenic zone. Segments in set D were recorded from the epileptogenic zone and those in set C from the hippocampal formation of the adverse hemisphere of the brain. While sets C and D comprised only activity estimated during seizure-free intervals, set E only included epileptic seizure activity. A summary of the datasets is shown in Table \ref{TBL:summary of datasets}.

\begin{table*}[!ht]
	\centering
	\caption{A summary of the clinical data.}
	\label{TBL:summary of datasets}
	\begin{tabular}{ L{4cm}L{2.4cm}L{2.4cm}L{2.4cm}L{2.4cm}L{2.4cm}  }
		\hline
		\textbf {Settings} &	\textbf {Set A} & \textbf{Set B} &	\textbf{Set C} &	\textbf{Set D} &	\textbf{Set E} \\  		\midrule
		Subjects &	5 healthy &	5 healthy &	5 epileptic patients &	5 epileptic patients &	5 epileptic patients \\ 
		
		Electrode type &	surface	 &	surface	 & Intracranial  & Intracranial &	Intracranial \\ 		
		
		Electrode placement	& International 10-20 system & International 10-20 system	&Hippocampal formation &	Epileptogenic zone &	Epileptogenic zone \\ 	
		
		Patient's state	& Awake, eyes open & Awake, eyes closed	 & Seizure-free (Interictal) &	Seizure-free (Interictal)	& Seizure activity (Ictal) \\ 	
		
		Number of epochs &	100 &	100 &	100 &	100 &	100
		\\  		
		Epoch duration (s)  &	23.6 &	23.6 &	23.6   &	23.6   &	23.6 \\ \hline
	\end{tabular}
\end{table*}

\subsection{Discrete Wavelet Transforms (DWT)} 
Wavelet transforms are widely used in many fields of engineering to solve many real-life problems. A wavelet is a short wave that has intensified energy over time to provide a tool for analyzing transient signals, non-stationary or variable phenomena over time. If a signal does not change much over time, we would call it a stationary signal. The Fourier transform can be easily applied to stationary signals and can get a good result. However, many signals such as EEG are non-stationary and transient signals; in such situation Fourier transform cannot be applied directly. But time-frequency methods can be used such as DWT \cite{C2_43_kumar2014epilepticfuzzy}.


DWT can expose signal details in time and frequency domain with precision. This makes it become a robust tool in biomedical engineering, especially in detecting epileptic seizures. In this thesis, DWT is used to analyze EEG signals in different frequency bands. The DWT decomposes a specific signal in detail and approximation coefficients at the first level. Then the coefficients of approximation are further subdivided into the next level of approximation and detail coefficients \cite{faust2015wavelet}.

In DWT, a wavelet called the mother wavelet $\psi (t)$ is the main controller of signal transformation, and a proper selection of such wavelet $\psi (t)$ strongly affects results. $\psi(t)$ can be represented by Equation \ref{EQ:1}:
\begin{equation}
\label{EQ:1}
\psi(t) = \frac{1}{\sqrt{a}}\psi(\frac{t-b}{a})
\end{equation}
Where $\psi$, $a$ and $b$ are indicated as the wavelet function, scaling and shifting parameters, respectively.

The wavelet transforms was classified into two types: Continuous wavelet transform (CWT) and Discrete wavelet transform (DWT). The CWT is defined as follows \cite{li2017classification}.

\begin{equation}
CWT(a,b)=\int_{-\infty}^{\infty} x(t) \frac{1}{\sqrt{|a|}}  \psi (\frac{t-b}{a}) dt                 \hspace{2cm}                                    
\end{equation}

Where $x(t)$ is a signal to be processed. If the scales and shifts parameters are transformed into powers of two, called dyadic scales and positions then the wavelet analysis will be extremely more efficient. Such analysis is obtained from the DWT which is illustrated as the following:
\begin{equation}
DWT(j,k)=\int_{-\infty}^{\infty} x(t) \frac{1}{\sqrt{|2^j|}}  \psi (\frac{t-2^jk }{2^j}) dt                  \hspace{2cm}   
\end{equation}

Where, $a$ and $b$ are replaced by $2^j$ and $ 2^{j}k$, respectively.

Frequency of main signal can be identified using WT coefficients, which makes it easy to get characteristics of signal $X(t)$ in both time and frequency domains. There are a couple of functions that DWT employs for analysis: the scaling and wavelet functions. Those functions reflect low-pass and high-pass filters. Two down samplers and two filters are involved in each step of DWT. High pass filter is devoted to extract details $(D_i)$ about signals (high resolution) by using the down-sampled outputs, while the low pass filter finds out approximations $(A_i)$ about signal.

\subsection{Support Vector Machine (SVM)} 
SVM is a powerful classifier in the field of biomedical science for the detection of abnormalities from biomedical
signals. SVM is an efficient classifier to classify two different sets of observations into their relevant class. It is capable to handle high dimensional and non-linear data excellently. On the basis of the structure of training data sets, it helps to predict the important characteristics of unknown testing data. As in this paper, to evaluate the performance of the proposed technique we are having four test cases with two different sets of class so we preferred this classifier for better accuracy results. SVM mechanism is based upon finding the best hyperplane that separates the data of two different classes of category. The structural design of the SVM depends on the following: first, the regularization parameter is used to manage the amount of allowed overlap between classes. Second, kernel functions of nonlinear SVMs are used for mapping of training data from an input space to a higher dimensional feature space.

All kernel functions like linear, polynomial, radial basis function and sigmoid having some free parameters are called hyper parameters. Suitable kernel function and parameters are required to train SVM classifier and usually obtained by the cross-validation technique.
In this paper, we have used the following kernel functions of SVM Classifier to analyze the performance of different test cases problems.
Radial basis kernel function with width $ \sigma$ \cite{andrew2000introduction}:
\begin{equation}
K(x,y) = \exp(- || x-y|| ^2 / 2 \sigma ^2) \hspace{2cm}     
\end{equation}
Where, K(x, y) is termed as the kernel function, which is built upon the dot product of two invariant x and y.

\subsection{K-Nearest Neighbor (KNN)} \label{subsubsec:K-NEAREST NEIGHBOR (KNN)}
K-Nearest Neighbor (KNN) classifier is simple and robust to even noisy and large training data set. It is also adaptive in nature because of using local information for prediction of unknown data. It performs the classification task on the basis of frequent class of its nearest neighbors in the feature space \cite{cover1967nearest}. It works to find a testing sample\textquotesingle s class by the majority class of the k nearest training samples.

\subsection{Na{\"\i}ve Bayes (NB)} 
Na{\"\i}ve Bayes is a simple and efficient statistical method, which is based on Bayes theorem \cite{wang2016detection}. NB is a simple technique for constructing classifiers models that assign class labels to problem instances, represented as vectors of feature values, where the class labels are drawn from some finite set. It assume that the value of a particular feature is independent of the value of any other feature, given the class variable.

\subsection{K-fold Cross-validation} 
Cross-validation is the statistical practice of partitioning a sample of data into subsets such that the analysis is initially performed on a single subset, while the other subset(s) are retained for subsequent use in confirming and validating the initial analysis. The initial subset of data is called the training set; the other subset(s) are called validation or testing sets \cite{C2_40_kumar2014epileptic}. In K-fold cross-validation, the original sample is partitioned into K sub-samples. K-1 sub-samples are used as training data, and single sub-sample is retained as the validation data for testing the model. The cross-validation process is then repeated K times, with each of the K sub-samples used exactly once as the validation data. The K results from the folds, then, can be averaged to produce a single estimation. In this study, we have used default 10-fold scheme to achieve best performance accuracies.

\subsection{Performance Evaluation Measurements} 
In this paper the set A, B, C and D are considered as positive class and set E is considered as the negative class respectively. To evaluate the classification performance for different test cases in this paper, we have used the using five measures, namely classification accuracy, sensitivity, specificity, precession and F Measure, and. The definitions of these measures are as follows: 
\begin{equation}
 \hspace{-1.5cm}Accuracy (Acc) =\frac{TP+TN}{TP+FN+TN+FP}*100     
\end{equation}
\begin{equation}
Sensitivity =\frac{TP}{TP+FN}*100 \hspace{5cm}     
\end{equation}
\begin{equation}
Specificity =\frac{TN}{TN+FP}*100 \hspace{5cm}     
\end{equation}
\begin{equation}
Precision=\frac{TP}{TP+FP}*100 \hspace{5cm}     
\end{equation}
\begin{equation}
F-Measure=2* \frac{Precision*Sensitivity}{Precision+Sensitivity} \hspace{2cm}     
\end{equation}
Where, True Positive (TP) stands for correctly identified non-seizure activity, True Negative (TN) is the correctly identified seizure activity, False Positive (FP) is the false identification of non-seizure activity, and False Negative (FN) is the falsely recognized seizure activity.

\section{The Proposed Classification Model}  
\label{model}

The proposed classification approach consists of three phases; namely, 1) pre-processing used to remove the noises from the EEG signals, 2) feature extraction used to extract the EEG signal features from decomposed signal, and 3) classification phases in this phase, the Extracted features are given as inputs to the classifier. The classification phase is mainly used to analyses the EEG signals and it classifies the EEG signal into normal or abnormal.

\begin{figure*}[!ht]
	\centering
	\includegraphics[width = 11cm,height=6cm]{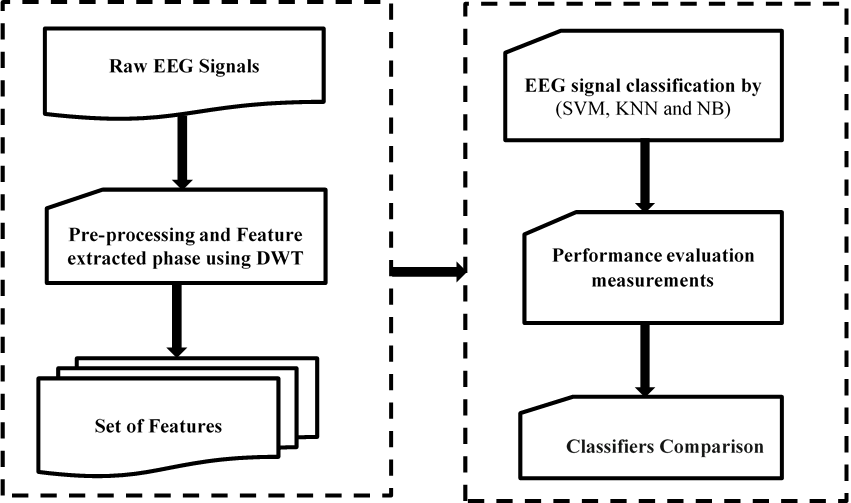}
	\caption{Block diagram of the proposed classification model.}
	\label{fig:mod}       
\end{figure*}

In the present work, EEG data sets (A, B, C, D and E) are preprocessed by DWT to decompose into five sub-band signals using four level decomposition. Next, useful features like Entropy, Min, Max, Mean, Median, Standard deviation, Variance, Skewness, Energy and Relative Wave Energy (RWE) are derived from each sub-band of wavelet coefficients. Finally, Extracted features are applied as input to SVMRBF, KNN and NB classifier for epilepsy classification. The block diagram of the proposed approach is shown in Figure \ref{fig:mod}.

\subsection{EEG Pre-processing}
\label{Sec :pre-processing}

For the EEG pre-processing phase, DWT decomposition has been used as a pre-processing level for EEG segments to extract five physiological EEG bands: delta (0-4 Hz), theta (4-8 Hz), alpha (8-12 Hz), beta (13-30 Hz), and gamma (30-60 Hz). In the first stage of the DWT, the signal is concurrently passed through an LP and HP filters. The outputs from low and high pass filters are indicated to as approximation (A1) and detailed (D1) coefficients of the first level. The output signals holding half the frequency bandwidth of the original signal can be downsampled by two due to the Nyquist rule. The same procedure can be duplicated for the first level approximation and the detailed coefficients fetch the second level coefficients. Through each step of this decomposition process, the frequency resolution is multiplied through filtering and the time resolution is split through down-sampling.

\begin{figure*}[!ht]
	\centering
	\includegraphics[width = 10cm,height=8cm]{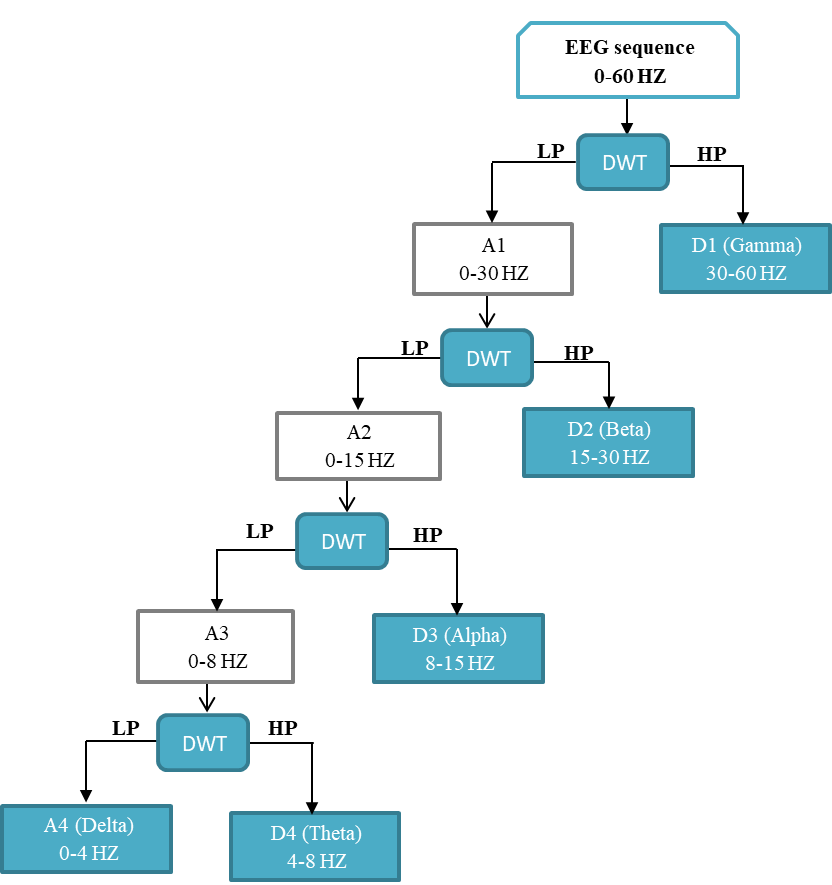}
	\caption{Four level wavelet decomposition of EEG.}
	\label{fig:Four level}       
\end{figure*}

Since the sampling frequency of the used EEG dataset is 173.61 Hz as shown in section \ref{Sec:EEG}, according to the Nyquist sampling theorem, the maximum useful frequency is half of the sampling frequency or 86.81 Hz. As such, from a physiological standpoint, frequencies greater than 60 Hz can be classified as noise and discarded. Consequently, to correlate the wavelet decomposition with the frequency ranges of the physiological sub-bands, the wavelet filter used in this application requires the frequency content to be limited to the 0–60 Hz band. Thus we have eliminated the frequencies above 60 Hz using a low-pass Butterworth filter. The band-limited EEG is then subjected to four-level DWT with fourth-order Daubechies (db4) wavelet function.  After the first level of decomposition, the EEG signal (0-60 Hz), is decomposed into its higher resolution components, D1 (30-60 Hz) and lower resolution components, A1(0-30 Hz). In the second level of decomposition, the A1 component is further decomposed into higher resolution components, D2 (15-30 Hz) and lower resolution components, A2 (0-15 Hz). Following this process, after four levels of decomposition, the components extracted are A4 (0-4 Hz), D4 (4-8 Hz), D3 (8-15 Hz), D2 (15-30 Hz), and D1 (30-60 Hz) as shown in Figure \ref{fig:Four level}. Reconstructions of these five components using DWT approximately correspond to the five physiological EEG sub-bands delta, theta, alpha, beta, and gamma.  Minor differences in the boundaries between the components compared to those between the EEG sub-bands are of little consequence due to the physiologically approximate nature of the sub-bands. The entire quantitative analysis of the EEG signals was coded using MATLAB (R2015a) and the Wavelet function. 

\subsection{Feature Extraction}\label{Sec :FE USING DWT}
Extracting the features consider the best depict of the behavior of EEG signals and are important for automated seizure detection performance. Feature extraction aims to capture the meaningful and distinctive characteristics hidden in EEG signals, which immediately dominates the final classification accuracy.
In this paper, we have extracted the following features of wavelet coefficients from each sub-band that were chosen to classify EEG signals \cite{panda2010classification, benzy2015combined}.
\\
1) Maximum of the wavelet coefficients in each sub-band.\\
2) Minimum of the wavelet coefficients in each sub-band.\\
3) Mean of the wavelet coefficients in each sub-band is obtained by the following Equation:
\begin{equation}
\mu _{i}=\frac{1}{N} \sum _{j=1}^{N} D_{ij}     \hspace{1cm}      i=1,2,...l       \hspace{2cm}                               
\end{equation}
4) The standard deviation of the wavelet coefficients in each sub-band.
The square root of the variance $\mu$ describes the mean value of the signal by the following equation.
\begin{equation}
\sigma=\sqrt{\frac {1}{N-1}  \sum_{i=1}^{N} (D_i-\mu)^2 }                       \hspace{2cm}                                    
\end{equation}
5) The variance of the wavelet coefficients in each sub-band is the square of the standard deviation.
\begin{equation}
V=\sigma^2
\end{equation}
6) The median of the wavelet coefficients in each sub-band.
The median of a statistical distribution D(x) is the value x such for a symmetric distribution; it is, therefore, equal to the mean. Given the statistical median of the random sample is defined by:

\begin{equation}
median=
\begin{cases}
D(\frac {N+1}{2} ),& \text if \hspace{0.25cm} N \hspace{0.25cm} is \hspace{0.25cm} odd \\\\
\frac {1}{2} (D(\frac {N}{2} )+D(\frac {N}{2} +1 )),              & \text if\hspace{0.25cm} N\hspace{0.25cm} is \hspace{0.25cm}even
\end{cases}
\end{equation}
7) Skewness of the wavelet coefficients in each sub-band.
A measure of the asymmetry of the data distribution. $\mu$ and $\sigma$ describe the mean and standard deviation of the signal individually by the following equation:
\begin{equation}
Skewness=
\frac {1}{N} \sum_{i=1}^{N}(\frac{D_i-\mu}{\sigma})^4 -3	                                             
\end{equation}
8)Energy in the sub-band
The energy points out that the strength of the signal as it gives the area under the curve of power at any interval of time. The energy of EEG signal of finite length is given by:
\begin{equation}
Energy (E_{i})=\sum_{j=1}^{N} |D_{ij}|^2  \hspace{1cm}      i=1,2,3...l	                            
\end{equation}
9) Relative Wave Energy (RWE) in the sub-band
RWE characterize the relative energy in each frequency sub-band and is utilize to detect the correspondence between segments of EEG signal. Energy of wavelet coefficients gives information about the strength of signals and is obtained by the equation:
\begin{equation}
E_j=\sum_{k}|D_{ik}|^2      \hspace{1cm}                 j=1,2,3...N                                  
\end{equation}
Where, j is the decomposition level and k is the corresponding wavelet coefficient. Moreover, total energy of decomposed levels of a signal segment is calculated by:
\begin{equation}
E_{Total}=\sum_j E_j     \hspace{1cm}     j=1,2,3...N 
\end{equation}
Relative wave Energy (RWE) is obtained by the equation:
\begin{equation}
\rho_{j}=E_j/E_{Total}        \hspace{1cm}       j=1,2,3...N   
\end{equation}
10) Entropy in the sub-band.
Entropy is a numerical measure of uncertainty (doubt) of outcome where signal contained thousands of bits of information. The mathematical representation is:
\begin{equation}
Entropy(EN)=\sum_{j=1}^{N} D_{ij}^2  \log (D_{ij}^2)   \hspace{1cm}    i=1,2,3...l               
\end{equation}	
Based on the above mentioned, ten features were extracted for all categories of signals to create the original feature database at each decomposition level starting from D1\textendash D4 and one final approximation, A4. These are extracted to help in distinguishing between normal and epileptic signal.

\subsection{Classification} 
The classification technique helps to discriminate the unknown testing set of observations into their appropriate
classes on the basis of the training set of known observations. A classification technique used a mathematical function named as a classifier to predict the right class of unknown observation of testing data set. In this paper, we have used three well-known supervised machine learning classification method named as SVMRBF classifier, KNN classifier and NB classifier for the evaluation of the performance of the proposed technique by utilizing the resulting features extracted from feature extraction technique.

\section{Experimental Results and Discussion}  
\label{Results}
\subsection{Experimental Results} 
\begin{figure*}[!htbp]
	\centering
	\includegraphics[width = 13cm,height=7cm]{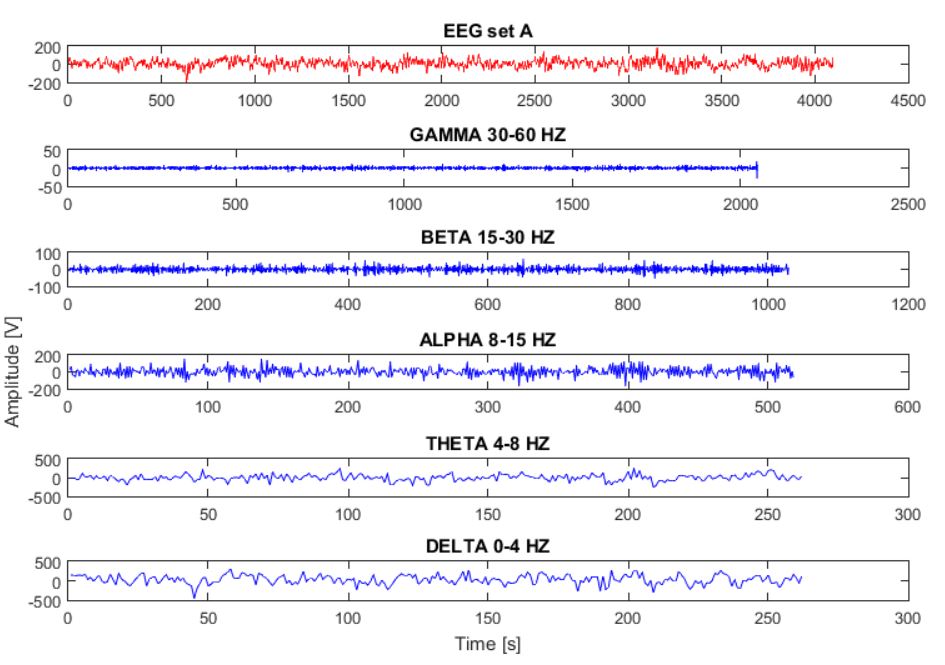}
	\caption{Approximate and detail coefficients for healthy subject (set A).}
	\label{fig: set A}       
\end{figure*}

\begin{figure*}[!htbp]
	\centering
	\includegraphics[width = 13cm,height=7cm]{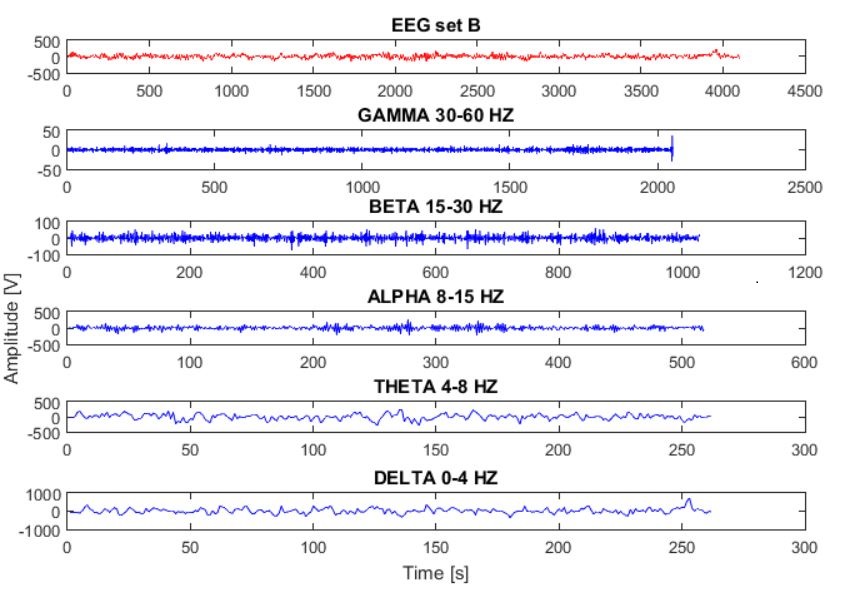}
	\caption{Approximate and detail coefficients for healthy subject (set B).}
	\label{fig: set B}       
\end{figure*}

\begin{figure*}[!htbp]
	\centering
	\includegraphics[width = 13cm,height=7cm]{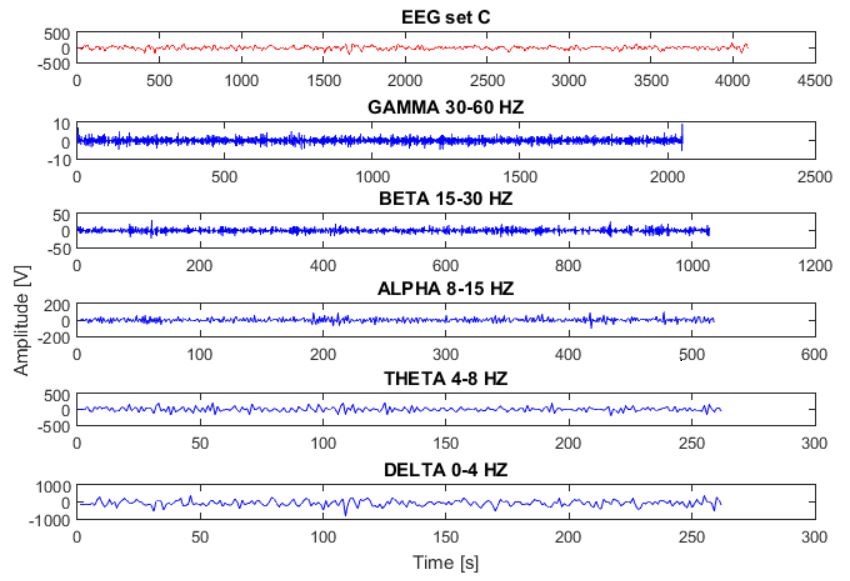}
	\caption{Approximate and detail coefficients for epileptic subject (set C).}
	\label{fig: set C}       
\end{figure*}

\begin{figure*}[!htbp]
	\centering
	\includegraphics[width = 13cm,height=7cm]{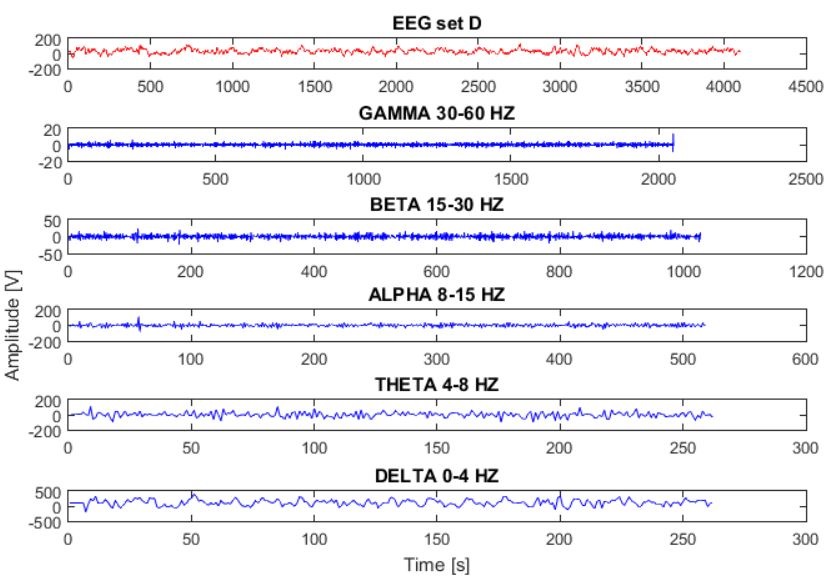}
	\caption{Approximate and detail coefficients for epileptic subject (set D).}
	\label{fig: set D}       
\end{figure*}

\begin{figure*}[!htbp]
	\centering
	\includegraphics[width = 13cm,height=7cm]{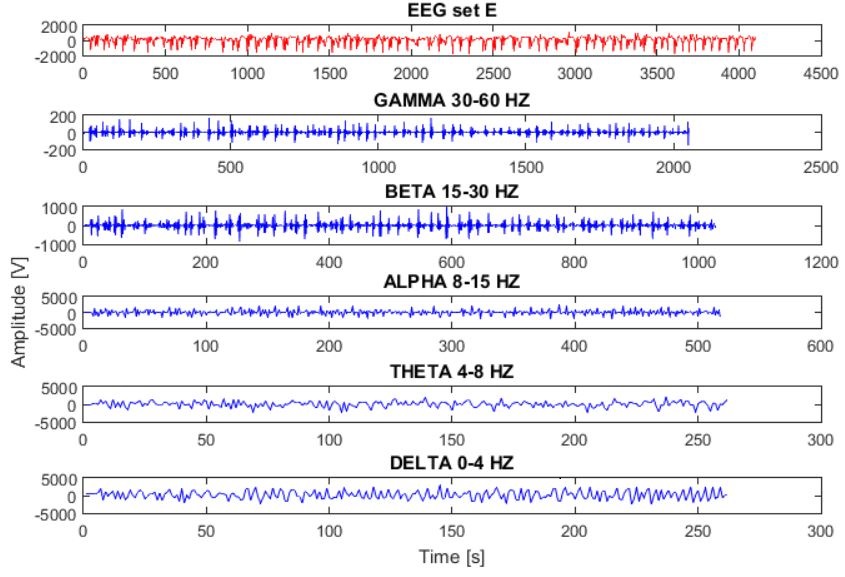}
	\caption{Approximate and detail coefficients for epileptic subject (set E).}
	\label{fig:set E}       
\end{figure*}

Extracting original features of Epileptic EEG, are done in two steps. In the first step, DWT is applied to decompose the EEG signal into several sub-signals within different frequency bands. Selection the number of decomposition levels and suitable wavelet function are also important for EEG signal analysis with DWT. In the current paper, the number of decomposition levels is chosen 4, which is recommended by others work \cite{C2_39_guo2011automatic}. And the wavelet function selected is Daubechies of order 4, which was also proven to be the best suitable wavelet function for epileptic EEG signal analysis \cite{C2_39_guo2011automatic}. The frequency bands responding to 4-level DWT decomposition with a sampling frequency of 173.6 Hz on the EEG signal are shown in Table {TBL:bands}. These Daubechies wavelet coefficients were computed and analyzed using MATLAB (R2015a). The five different sub-signals (one approximation A4 and four details D1\textendash D4 that correspond to delta (0-4 Hz), gamma (30-60 Hz), beta (13-30 Hz), alpha (8-12 Hz), and theta (4-8 Hz) respectively, of the sample  EEG epoch taken from data sets A,B,C,D and E are plotted in Figures \ref{fig: set A}  to \ref{fig:set E} respectively. 

The second step, after raw EEG signal, is decomposed into five sub-signals, which individually correspond to different frequency bands described in Table \ref{TBL:bands}. Ten classic Features explained in Section \ref{Sec :FE USING DWT} are calculated, using MATLAB (R2015a), from the approximation and detail coefficients of all sub-bands of the entire 500 EEG epochs of five data sets A\textendash E to form the original feature database. Theses extracted features from each sub-band for the last epoch of data sets A, D and E for instance are presented in \cite{hamad2016feature}. 

\begin{table}[htbp]
	\centering
	\caption{Frequency bands of EEG signal with 4-Level DWT decomposition.}
	\label{TBL:bands}
	\begin{tabular}{lll}
		\midrule
		\textbf{level} & \textbf{Sub-band signal} & \textbf{Frequency band (Hz)}
		\\ 	\midrule
		1                   & D1(gamma)       & 30\textendash60                                                             \\ 	
		2                   & D2 (beta)       & 15-30                                                             \\ 	
		3                   & D3 (alpha)      & 8-15                                                             \\ 	
		4                   & D4 (theta)      & 4-8                                                               \\ 	
		4                   & A4 (delta)      & 0-4                                                              \\    \midrule
	\end{tabular}
\end{table}

\subsection{Discussion} 
The features of D1\textendash D4 and A4 that are extracted in Section \ref{Results} were classified using SVMRBF, KNN and NB. These features are used as an input of classifiers to classify the EEGs as healthy, interictal and ictal. The proposed approach is tested on the four different test cases. The SVMRBF, KNN and NB are implemented by using MATLAB (R2015a). 

The input feature vector is randomly divided into training data set and testing data set based 10-fold cross-validation. The training data set is used to train theses classifiers, whereas the testing data set is used to verify the accuracy and effectiveness of the trained classifiers for the given EEG classification problem. Each row of the input data matrix is one observation and its column is one feature. In this work, the feature vector of data set A has 100 rows and 50 columns. Similarly, the feature vector of sets B, C, D and E individually have 100 observations and 50 features. The data set for the present binary classifier task consists of 200 observations of 50 features for case 1 to case 4 as shown in Figure \ref{TBL:cases}. The training data set consists of 90\% of input data and the remaining 10\% of input data are used for testing of the classifiers. This process is repeated 10 times to obtain the average values of statistical parameters which are summarized in Table \ref{TBL:svm} to \ref{TBL:NB} of SVMRBF, KNN and NB classifiers respectively.


\begin{table*}[!htpb]
	\centering
	\caption{The classification description of different test cases along with their EEG data sets.}
	\begin{tabular}{{L{3cm}L{3cm}L{7cm}}}
		\hline
		\textbf{\begin{tabular}[c]{@{}l@{}}Test  case\end{tabular}} & \multicolumn{1}{l}{\textbf{\begin{tabular}[c]{@{}c@{}}Cases for seizure \end{tabular}}} & \multicolumn{1}{l}{\textbf{\begin{tabular}[c]{@{}c@{}}Classification description\end{tabular}}} \\ \hline
		\textbf{Case 1}                                                
		& Set A vs Set E                                                                                      & Healthy Persons with eye open vs \\
		&&Epileptic patients during seizure activity                                 \\ 
		\textbf{Case 2}                        
		& Set B vs Set E                                                                                      & Healthy Persons with eye close vs \\
		&&Epileptic patients during seizure activity                                \\ 
		\textbf{Case 3}                   
		& Set C vs Set E                                                                                      & Hippocampal seizure free vs \\
		&&Epileptic patients during seizure activity                                      \\
		\textbf{Case 4}                    
		& Set D vs Set E                                                                                      & Epileptic seizure free vs \\
		&&Epileptic patients during seizure activity                                        \\ \hline
	\end{tabular}
	\label{TBL:cases}
\end{table*}

\begin{table*}[!htbp]
	\centering
	\caption{The performance for different sets of EEG data using SVMRBF.}
	\label{TBL:svm}
	\begin{tabular}{llllll}
		\hline
		\multicolumn{1}{c}{\textbf{\begin{tabular}[c]{@{}c@{}}Cases for seizure \end{tabular}}} & \multicolumn{5}{c}{\textbf{SVMRBF}}                           \\ \hline
		\textbf{}                                                    & Accuracy(\%) & Sensitivity(\%) & Specificity(\%) & Precision (\%) & F-Measure(\%) \\ \hline
		\textbf{Set A vs Set E}                                                                               & 100           & 100              & 100              & 100            & 100           \\ 
		\textbf{Set B vs Set E}                                                                               & 100           & 100              & 100              & 100            & 100           \\ 
		\textbf{Set C vs Set E}                                                                               & 99            & 100              & 98               & 98.039         & 99.01         \\ 
		\textbf{Set D vs Set E}                                                                                        & 97            & 98               & 96               & 96.078         & 97.03         \\ \hline
	\end{tabular}
\end{table*}

\begin{table*}[!htpb]
	\centering
	\caption{ The performance for different sets of EEG data using KNN.}
	\label{TBL:knn}
	\begin{tabular}{llllll}
		\hline
		\multicolumn{1}{c}{\textbf{\begin{tabular}[c]{@{}c@{}}Cases for seizure \end{tabular}}} & \multicolumn{5}{c}{\textbf{KNN}}                                                    \\ \hline
		\textbf{}                                                                                             & Accuracy(\%) & Sensitivity(\%) & Specificity(\%) & Precision (\%) & F-Measure(\%) \\ \hline
		\textbf{Set A vs Set E}                                                                               & 99.5          & 99               & 100              & 100            & 99.497        \\ 
		\textbf{Set B vs Set E}                                                                               & 99            & 98               & 100              & 100            & 98.99         \\ 
		\textbf{Set C vs Set E}                                                                               & 97.5          & 95               & 100              & 100            & 97.436        \\ 
		\textbf{Set D vs Set E}                                                                               & 96.5          & 94               & 99               & 98.947         & 96.41         \\ \hline
	\end{tabular}
\end{table*}

\begin{table*}[!htbp]
	\centering
	\caption{The performance for different sets of EEG data using NB.}
	\label{TBL:NB}
	\begin{tabular}{llllll}
		\hline
		\multicolumn{1}{c}{\textbf{\begin{tabular}[c]{@{}c@{}}Cases for seizure \end{tabular}}} & \multicolumn{5}{c}{\textbf{NB}}                                                     \\ \hline
		\textbf{}                                                                                             & Accuracy(\%) & Sensitivity(\%) & Specificity(\%) & Precision (\%) & F-Measure(\%) \\ \hline
		\textbf{Set A vs Set E}                                                                               & 99.5          & 100              & 99               & 99.01          & 99.502        \\ 
		\textbf{Set B vs Set E}                                                                               & 99            & 99               & 99               & 99             & 99            \\ 
		\textbf{Set C vs Set E}                                                                               & 98.5          & 99               & 98               & 98.02          & 98.507        \\ 
		\textbf{Set D vs Set E}                                                                               & 96.5          & 95               & 98               & 97.938         & 96.447        \\ \hline
	\end{tabular}
\end{table*}

\subsection{Comparison Analysis} 
There are many other methods proposed by different researchers for the epileptic seizure detection. Table \ref{com} presents a comparison of the results between the method developed in this work and other methods proposed in the literature in terms of accuracy. Only methods evaluated on the same dataset for the same cases are included so that a comparison between the results is feasible.

\section{Conclusion and Future Work}
\label{Sec:Conclusion}
The detection of epileptic seizure being performed by visual scanning of EEG signal is very time-consuming, costly procedure and may be inaccurate, specifically for a long time EEG recording. In this paper the DWT is used for analysis of EEG to detect epilepsy. EEG signals are decomposed into different sub-bands through DWT to obtain the detail wavelet coefficients (D1\textendash D4) and approximate wavelet coefficients (A4). The sub-band coding gives different frequency bands which are Gamma (D1: 30-60 Hz), Beta (D2: 15-30 Hz), Alpha (D3: 8-15 Hz), Theta (D4: 4-8 Hz), and Delta (A4: 0-4 Hz). Then, ten features were extracted using DWT from  each sub-band to classify EEG signal. Furthermore,  three different classifiers (SVMRBF, KNN and NB) were employed and their performance was evaluated for distinguishing between normal and epileptic. 
The best classification accuracies are obtained using SVMRBF for cases 1 and 2 is 100\%. Finally, the proposed method is verified by comparing the performance of classification problems as addressed by other researchers. It can be concluded that using DWT based proposed features; more satisfactory results are achieved to discriminate the EEG signals in comparison to other methods. The proposed method can be employed as a quantitative measure for monitoring the EEG and it may prove to be a useful tool in analyzing the EEG signal associated with epilepsy. As future work, the proposed approach can be applied to a more wide range of pattern recognition problems which are important to humans, such as the Alzheim’s and Parkinson’s diseases detection and diagnosis. 

\section*{References}
\bibliography{References}

\begin{table}[h]
	
	\centering
	\caption{
		A comparison of classification accuracy obtained by our method and others’ method for binary EEG classification problem.}
	\label{com}
	\begin{tabular}{L{0.8cm}L{1.5cm}L{5cm} L{0.9cm}}
		
		\hline
		\textbf{Cases} & \textbf{Ref}
		& \multicolumn{1}{c}{\textbf{Methods}} 
		& \textbf{Acc(\%)} \\ \hline
		
	\multirow{6}{*}{\textbf{\begin{tabular}[c]{@{}l@{}}A vs \\ E \end{tabular}}}  & {\cite{C2_39_guo2011automatic}}                                                       & \begin{tabular}[c]{@{}l@{}}Genetic programming-based \\KNN classifier\end{tabular}                         & 99.2                 
		\\  
		
		& {\cite{C2_42_wang2011best}}                                                      & Wavelet packet entropy with KNN
		& 99.449         
		\\ 
		& {\cite{C2_41_li2011clustering}}                                                      &  clustering technique-based least square support vector machine (CT-LS-SVM)                                                                                                                          & 99.90                  \\

		& {\cite{C2_46_guo2010epileptic}}                                                       & Multiwavelet transform based  approximate entropy feature with artificial neural networks. & 99.85                              \\ 
		& {\cite{C2_36_nicolaou2012detection}}                                                      & Permutation Entropy with SVM                                                                                                         & 93.55         \\

			& \textbf{\begin{tabular}[c]{@{}l@{}}Proposed \\ \end{tabular}} & \textbf{SVMRBF, KNN and NB}                                         & \textbf{100 99.5 99.5}      
		   \\ \hline

		\multirow{6}{*}{\textbf{\begin{tabular}[c]{@{}l@{}}B vs \\ E \end{tabular}}}

		& {\cite{C2_40_kumar2014epileptic}}                                                      & DWT based approximate entropy (ApEn) with Artificial neural network  & 92.5              \\
		&{\cite{C2_41_li2011clustering}}                                                      &  clustering technique-based least square support vector machine (CT-LS-SVM)                                                                                                                         & 96.30                 \\ 
		
		& {\cite{C2_45_supriya2016weighted}}                                                      & Weighted Visibility Graph with SVM   & 97.25 \\               	& {\cite{C2_36_nicolaou2012detection}}                                                      & Permutation Entropy with SVM                                                                                                         & 82.88        \\

		&\textbf{\begin{tabular}[c]{@{}l@{}}Proposed \\ \end{tabular}} 
	& \textbf{SVMRBF, KNN and NB}
                               
	&\textbf{100 99 99}            \\ \hline

		\multirow{6}{*}{\textbf{\begin{tabular}[c]{@{}l@{}}C vs \\ E \end{tabular}}} 
		&{\cite{C2_41_li2011clustering} }                                                     &  clustering technique-based least square support vector machine (CT-LS-SVM)                                                                                                                 &96.20             \\
		&   {\cite{C2_45_supriya2016weighted}}                                                      & Weighted Visibility Graph with SVM  & 98.25 \\               	& {\cite{C2_36_nicolaou2012detection}}                                                      & Permutation Entropy with SVM                                                                                                         & 88.83        \\

		& \textbf{\begin{tabular}[c]{@{}l@{}}Proposed \\ \end{tabular}} & \textbf{SVMRBF, KNN and NB}                                                                                                                                  & \textbf{99 97.5 98.5}          \\ \hline

	\multirow{6}{*}{\textbf{\begin{tabular}[c]{@{}l@{}}D vs \\ E \end{tabular}}} 
		& {\cite{C2_40_kumar2014epileptic}}                                                      & DWT based approximate entropy (ApEn) with Artificial neural network  & 95                     \\
		& {\cite{C2_43_kumar2014epilepticfuzzy}}                                                      & DWT based fuzzy approximate entropy and SVM
		& 95.85                  \\  
		&{\cite{C2_41_li2011clustering}  }                                                    &  clustering technique-based least square support vector machine (CT-LS-SVM)                                                                                                                         & 93.60                    \\ 
		
		&   {\cite{C2_45_supriya2016weighted}}                                                      & Weighted Visibility Graph with SVM  & 93.25 \\               	& {\cite{C2_36_nicolaou2012detection}}                                                      & Permutation Entropy with SVM                                                                                                         & 83.13       \\

		& \textbf{\begin{tabular}[c]{@{}l@{}}Proposed \\ \end{tabular}} & \textbf{SVMRBF, KNN and NB}                                                                                                                             & \textbf{97 96.5 96.5}          \\ \hline
	\end{tabular}
\end{table}

\end{document}